\documentstyle[12pt,epsfig]{article}
\begin{document}
\begin{center}
{\Large{\bf{Study of ATLAS  Sensitivity to FCNC Top Quark
 Decay
$t \rightarrow Zq$}}}
\end{center}
\   \par
\   \par
\begin{center}
{\large{\bf{Leila Chikovani}}}
\end{center}
\begin{center}
Institute of Physics of the Georgian Academy of Sciencies,\par
\end{center}
\   \par
\   \par
\begin{center}
{\large{\bf{Tamar Djobava}}}
\end{center}
\begin{center}
High Energy Physics Institute, Tbilisi State University,\par
\end{center}
\  \par
\  \par
\begin{center}
{\bf{ABSTRACT}}
\end{center}
\  \par
\  \par
The sensitivity of ATLAS experiment to the
top-quark rare decay via flavor-changing neutral currents
$t \rightarrow Zq$ ($q$ represents $c$ and $u$ quarks) have been studied
at $\sqrt{s}$=14 TeV in two
decay modes:
1.The pure leptonic decay of gauge bosons:
$t\bar{t} \rightarrow ZqWb \rightarrow l^{+} l^{-} j l^{\pm} \nu j_{b}$ ,
(l=e, $\mu$).
2.The leptonic decay of Z bosons and hadronic decay of W bosons:
$t\bar{t} \rightarrow ZqWb \rightarrow l^{+} l^{-} jjjj_{b}$ ,
(l=e, $\mu$).
 The dominant backgrounds
 Z+jets, ZW and $t\bar{t}$ has been analysed. The signal
and backgrounds were ganarated via PYTHIA 5.7, simulated  and analysed using
ATLFAST 2.14.
 A branching ratio for $t \rightarrow Zq$ as low
as 1.1x10$^{-4}$ for the leptonic mode
and  2.3x10$^{-4}$ for hadronic mode
could be discovered at the 5$\sigma$ level with an integrated luminosity
of 10$^5$ pb$^-1$.
\newpage
{\Large{\bf{1.Introduction}}}
\   \par
\   \par
\   \par
\par
The existence of the top quark has been established at the Fermilab
Tevatron by the CDF and DO Collaborations [1]. Because the top quark is heavier
than all other observed fermions and gauge bosons and has a mass of the
order of the Fermi scale, it couples to electroweak symmetry breaking sector
strongly. If anomalous top quark couplings beyond the Standard Model (SM)
were to exist, they would affect top quark production and decay processes
at hadron and $e^{+}$ $e^{-}$ colliders [2,3].
Therefore to study the flavour-changing neutral current top quark
decay $t \rightarrow Zq$ ($q$ represents either $c$ or $u$ quarks)
is of great interest, as known [4,5] that LHC can be considered as
"top factory", because about 80 000
$t\bar{t}$ events  are expected to be produced per day   at $L=10^{33}~
cm^{-2}~s^{-1}$, that's why
the LHC with high
integrated luminosity and energy have good potential to explore this rare
decay.
\par
We study the sensitivity of ATLAS experiment to the Branching ratio
of top-quark rare decay mode $t \rightarrow Zq$ ($q=u,c$).
In the framework of the Standard Model
the loop suppression and heaviness of gauge bosons makes this process
rare being of an order of 10$^{-13}$ [6]. Meanwhile the other models predict
significantly larger Branching ratios for this process. Two-Higgs-doublet
model predicts for Br($t \rightarrow Zq$)$\sim 10^{-9}$ [7]
and Super symmetric
(SUSY) model (without $R$-parity) - Br($t \rightarrow Zq$)$\sim 10^{-4}$
[8]. The existing limit at 95 $\%$ CL from CDF [9] is
Br($t \rightarrow Zq$) $< 33\%$ (RUN1 at the Tevatron).
The observation of such
a top-quark decay mode would signal to existance of new physics - the
physics beyond the SM: new dynamical interactions of top quark,
multi-Higgs doublets, exotic fermions and etc. [5,6,7]
In addition this mode is of great interest since
$t\bar{t} \rightarrow ZZ + cc$ decays would prove to be serious background
to events containing $Z$ boson pairs and jets from cascade decays of
squarks and gluinos [10].
\par
The dominant mechanisms for top quarks production at LHC are
$q\bar{q}$ and $gg$ annihilation channels
$q\bar{q}$, $gg \rightarrow t\bar{t}$.
The final state topology of $t\bar{t} \rightarrow ZqWb$ has two decay mode
of gauge bosons:\\
1. The pure leptonic decay of gauge bosons:\\
$t \rightarrow Zq \rightarrow l^{+} l^{-} j$ ,
$\bar{t} \rightarrow Wb \rightarrow l^{\pm} \nu j_{b}$ (l=e, $\mu$)\\
2. The leptonic decay of Z bosons and hadronic decay of W bosons:\\
$t \rightarrow Zq \rightarrow l^{+} l^{-} j$ ,
$\bar{t} \rightarrow Wb \rightarrow jjj_{b}$ (l=e, $\mu$)\\
The value of the Branching ratio of hadronic mode is three times more than
the leptonic one, but it has enormous QCD-backgrounds
at hadron collider while the pure leptonic has a very distinct experimental
signature: three isolated
charged  leptons, two of which reconstruct a $Z$, a large missing
transverse momentum ( $P_{T}^{miss}$ ) and two hard jets coming from the $b$
and $q$ ($q= c,u$) quarks.
\par
The both channels are of great interest.
\   \par
\   \par
\   \par
{\Large{\bf{2.The Signal Generation}}}
\   \par
\   \par
\   \par
\par
As was mentioned above
the dominant source of top quark production at hadron colliders are
$q\bar{q}$, $gg \rightarrow t\bar{t}$ processes.
The PYTHIA 5.7 was set up to produce $t\bar{t}$ events
at $\sqrt{s}=14$ TeV and $m_{top}=174$ GeV with proton structure
functions CTEQ2L. Initial and final state QED and QCD (ISR, FSR) radiation,
multiple interactions, fragmentations and decays of unstabled particles
were enabled, but the underlying event (beam remnants, pile up etc)
was switched off.
\par
The generation of the channel $t \rightarrow Zq$ is not implemented in the
PYTHIA. This process had been included into PYTHIA in the following way:
all individual decay channels of top quark had been switched off except of the
$t \rightarrow Wb$ and $t \rightarrow Ws$ ones. In the channel $t \rightarrow Ws$
the decay $t \rightarrow Zq$ was included replacing  $W$ by $Z$ and $s$ by
$c(u)$.
\   \par
\   \par
\   \par
{\Large{\bf{3.The Leptonic Decay Mode}}}
\   \par
\   \par
\   \par
\par
The signature for the pure leptonic decay mode as mentioned above is
$t\bar{t} \rightarrow ZqWb \rightarrow l^{+}l^{-}jl^{\pm}j_{b}$.
Due to the requirement of three isolated charged leptons, two of which reconstruct a
$Z$ and a large missing transverse energy
 the following SM backgrounds has been considered:\\
$\bullet$  $Z( \rightarrow ll) +jets$\\
$\bullet$ $pp \rightarrow W^{\pm}Z+X \rightarrow l^{\pm} \nu l^{+} l^{-}+X$\\
$\bullet$ $t\bar{t} \rightarrow W^{+}bW^{-} \tilde {b}
 \rightarrow l^{+} \nu j_{b}
l^{-} \tilde {\nu}j_{b}$.\\
\par
 The main goal is
to analyse fully generated events and select the isolated leptons, reconstruct
jets, identify b-jets and estimate the missing transverse  energy.
A parametrization of electron, muon momentum energy resolution is included, as
well as a parametrization of the hadronic calorimeter energy resolution
(for jets) and the effect of the ATLAS magnetic field on jet reconstruction.
For this purpose some subroutines from the fast simulation package
ATLFAST 1.0 have been used
[11]. For the leptons isolation only  tracking isolation method had been
considered. The jets momenta had been recalibrated by the calibration
factor
$K_{jet}=P_{T}^{parton}/P_{T}^{jet}$ according to the algorithm of the ATLFAST
package, where $P_{T}^{parton}$ denotes the transverse momentum $P_{T}$ of
the parton
which initiated the jet (before FSR).
\par
Some preselection cuts were applied to signal and backgrounds before
writing the events on DST:\\
1. The event contains at least 3 leptons (electrons with $P_{T} > 5 $ GeV
and muons with
$P_{T} > 6 $ GeV) within pseudorapidity $|\eta| < 2.5$.\\
2. The events must contain a pair of opposite sign and same flavour
leptons compatible with them having come from Z bosons.\\
3. The number of jets in the event $N_{jet} \geq 2$ with $P_{T}^{jet} > 15 $ GeV.
\par
The $Z( \rightarrow ll)+jets$ background in the
final state also have a pair of isolated charged
leptons, but $P_{T}^{miss}$ in such processes arises mainly from
 neutrinos produced in semileptonic decays of heavy quarks.
\par
$Z( \rightarrow ll)+jets$ production at LHC has relatively large cross section
dominated by $qg \rightarrow Zq $ and $q\bar{q} \rightarrow Zg $
processes. To speed up the generation, the thresholds on invariant mass
$\hat{m}=\sqrt{\hat{s}}$
and transverse momentum $\hat{p_{\perp}}$ had been imposed:
$\hat{m}=\sqrt{\hat{s}} > 130 $ GeV; $\hat{p_{\perp}} > 50$ GeV.
So we collect events mainly from the top peak (otherwise we would
have additional large contribution from undesireble events).
The cross section
for this sample of events is $\sigma_{Z+jets}$=3186 pb.
The demand  of
three isolated leptons and large P$_{T}^{miss}$ reduces this background
significantly.
\par
The $ZW$ background  is the electroweak
process
 $pp \rightarrow W^{\pm}Z+X \rightarrow l^{\pm} \nu l^{+} l^{-}+X$.
$\sigma_{ZW}$=26.58 pb.
Since the signal events naturally contain two energetic jets from heavy top
quark decays, typically with a transverse momentum of order
$P_{T}(j)\simeq 1/2m_{t}(1-M_{w}^{2}/m_{t}^{2})$, [5]
to demand of two observable jets in the events one of which is $b$-jet,
reduces the $WZ$ background significantly.
\par
In the selected $t\bar{t}$ background events sources of leptons are arising from
the decays of $W^{\pm}$ and from cascade decays, where quarks initially decay
hadronically, but produce leptons further in the decay chain.
These cascade leptons have a softer $P_{T}$ spectrum than the leptons from
$W$ and requirement of three isolated leptons with high $P_{T}$ reduce
$t\bar{t}$ background
significantly. We have to emphasize, that
$t\bar{t}$ background suffers from the demand of the existence only one tagged
$b$- jet in the event.
\par
The number of events collected over a given period of time is calculated
by the formula:
\begin{eqnarray}
N=\sigma \cdot L   \label{eq1}
\end{eqnarray}
where $L$ is the integrated luminosity ($L=10^{5} ~ pb^{-1}$). The expected
number of events $N_{exp}$ are:\\
$\bullet$ for the signal
\begin{eqnarray}
N_{exp}^{S}= 2 \cdot \sigma_{t\bar{t}} \cdot Br(t \rightarrow Zc) \cdot
Br(t \rightarrow Wb) \cdot Br(W \rightarrow l \nu) \cdot
Br(Z \rightarrow ll) \cdot L  \label{eq2}
\end{eqnarray}
\begin{center}
$N_{exp}^{S}= 22.8 \cdot Br(t \rightarrow Zc) \cdot L$
\end{center}
the estimation has been done for $\sigma_{t\bar{t}}$=800 pb [12].\\
$\bullet$ for $Z (\rightarrow ll)+jets $ background
\begin{center}
$N_{exp}^{Z+jets}= 21400000 $
\end{center}
$\bullet$ for $ZW$ background
\begin{center}
$N_{exp}^{ZW}= 40000$
\end{center}
$\bullet$ for $t\bar{t}$ background
\begin{eqnarray}
N_{exp}^{t\bar{t}}= \sigma_{t\bar{t}} \cdot
 (Br(W \rightarrow l \nu))^{2} \cdot
(Br(t \rightarrow Wb))^{2} \cdot L  \label{eq3}
\end{eqnarray}
\begin{center}
$N_{exp}^{t\bar{t}}= 3720000$
\end{center}
\par
Two sets of kinematical cuts had been applied in sequence for the signal
and backgrounds (see Tables 1$\div$9).
\  \par
\  \par
\begin{tabular}{lll}
&  S E T 1 & S E T 2\\
&  & \\
Lepton isolation cut:& ${\Delta}R= 0.2$  & ${\Delta}R= 0.3$\\
Lepton $P_{T}$ cut:&\multicolumn{2}{l}
{3l with $P_{T}^{l} > 20~ GeV$ in $|\eta^{l}| < 2.5 $}\\
$E_{T}$ missing:&\multicolumn{2}{l}{$E_{T}^{miss} > 30 ~GeV$} \\
jets $\eta$ cut:&\multicolumn{2}{l}{jets with $|\eta^{jet}| < 2.5 $}\\
2 jets $P_{T}$ cut:&\multicolumn{2}{l}{2jets with $P_{T}^{jet} > 40~ GeV$}\\
                   &\multicolumn{2}{l}{2jets with $P_{T}^{jet} > 50~ GeV$}\\
jets isolation cut:&\multicolumn{2}{l}{${\Delta}R_{jj} > 0.4 $}\\
Lepton-jets isolation cut:&\multicolumn{2}{l}{${\Delta}R_{lj} > 0.4 $}\\
Z mass cut:&\multicolumn{2}{l}{$M_{Z} \pm 6~ GeV$}\\
$b$ jets cut:&\multicolumn{2}{l}{one tagged $b$ jet in the event}\\
Top mass cut:&\multicolumn{2}{l}{$M_{Zq} \pm 8~ GeV$,
$M_{Zq} \pm 12~GeV$, $M_{Zq} \pm 24~GeV$}
\end{tabular}
\   \par
\   \par
For the lepton tracker (tracking)
isolation it had been assumed that there be no track with $P_{T} > 2~ GeV$ in cone
${\Delta}R=\sqrt{(\Delta\phi)^2+(\Delta\eta)^2}$. It have been considered
two cones of ${\Delta}R$
: ${\Delta}R= 0.2$ (SET1) and ${\Delta}R= 0.3$ (SET2).
\par
The requirement of 3 leptons on the level of preselection cuts reduces
significantly Z+jets and $t\bar{t}$ backgrounds, meanwhile the requirement
of two jets reduces significantly ZW and Z+jets backgrounds.
\par
The efficiency of preselection cuts for signal is $80\%$, while for the
large number expected events Z+jets background is $1.7\%$. After applying the
preselection cuts other backgrounds are rejected (ZW background by $\sim 90\%$
and $t\bar{t}$ by $\sim 80\%$) significantly.
\par
In Tables 1$\div$8 three leptons identification efficiency ((0.9)$^{3}$= 0.729)
have not been taken into consideration, but for the estimation of the
Branching ratios it have been done.
\par
Three leptons isolation cut reduces signal by $\sim$ 15 $\%$ for SET1 and
by $\sim$ 23 $\%$ for SET2. Thus for SET2 the signal reduction is more
 by $\sim$ 10 $\%$ than for SET1. The ZW background is not affected
by lepton isolation cut. This background reduces only by
$\sim$ 3$\div$4 $\%$ for both SETs. In Z+jets and  $t\bar{t}$ backgrounds
two islolated leptons are present and the third lepton originates from the
cascade decays. The requirement of three isolated leptons affects most significantly
these backgrounds, about 90 $\div$ 95 $\%$ of events respectively for set1
and set2 are rejected.
\par
The requirement of three leptons with
 $P_{T}^{l} > 20~ GeV$ in $|\eta^{l}| < 2.5 $ reduces signal and ZW
background by about 30 $\%$. In Z+jets and $t\bar{t}$ backgrounds
this cut remains about 4 $\%$ of events. This is well illustrated
in Figs.1 and 2. Fig.1 presents the $P_{T}$ distribution of two leptons
that reconstruct Z invariant mass for signal and backgrounds and Fig.2
shows the $P_{T}$ distribution of third lepton respectively. The arrows in Figs. 1
and 2 indicate the threshold $P_{T}$ for leptons $P_{T}^{l} = 20~ GeV$.
\par
After applying $P_{T}^{miss} > 30 ~ GeV$  cut in signal, ZW and $t\bar{t}$ backgrounds
remain 76 $\%$, 70 $\%$ and 90 $\%$ of events, which were accepted by the previous
cut, whereas for Z+jets remain only $\sim$ 8 $\%$ of events. Thus the
$P_{T}^{miss}$  cut  reduces significantly Z+jets background (see Fig.3.,
arrow indicates the threshold $P_{T}^{miss} = 30 ~ GeV$).
\par
It have been demanded the presence of two jets in $|\eta^{jet}| < 2.5 $
satisfying to the following conditions of isolation: ${\Delta}R_{jj} > 0.4 $
(jet-jets isolation) and ${\Delta}R_{lj} > 0.4 $ (lepton-jets isolation).
Two threshold values for $P_{T}^{jet}$ cut have been examined:
$P_{T}^{jet} > 40~ GeV$ and $P_{T}^{jet} > 50~ GeV$ (see Tables 1$\div$9).
Fig.4 presents the number
of jets with $P_{T}^{jet} > 50~ GeV$ in an event. One can see, that the cut requiring
the presence of 2 and more jets in each event is a major rejector for the
ZW background.
\par
The kinamatic
variable which can be used to seperate signal from backgrounds is the
reconstructed Z mass (since there is no Z in the $t\bar{t}$ background). For the
reconstruction of Z mass, have been required a pair of isolated leptons within
$M_{Z} \pm 6~ GeV$ mass window. The choice of 6 GeV for this cut is motivated mainly
by the resolution ($\sim ~2\sigma$) on the reconstructed Z in the signal.
Fig 5.a presents distribution
of reconstructed invariant mass of $ll$ pairs $m_{ll}$, for all combinations of
$ll$ for the signal. The Z mass cut does not affect strongly
the ZW and Z+jets backgrounds .
Meanwhile this cut is very effective tool for the suppression of $t\bar{t}$ background.
After applying this cut only 10$\div$12 $\%$ of events accpeted by previous
cut retain  for SET1 and $P_{T}^{jet} > 40~ GeV$  and  only 5$\div$7 $\%$ of events
for SET2 and $P_{T}^{jet} > 50~ GeV$.
\par
The next demand is the presence in the event only one tagged  $b$ jet.
After this cut in signal survive 50 $\%$ of events accepted by Z mass cut
for both SETs and $P_{T}^{jet}$  criteria whereas for $t\bar{t}$ background
this value is smaller by about 10$\div$20 $\%$.
For ZW background the one tagged
$b$ jet cut is effective rejector, after applying this cut survive
7$\div$8 $\%$ of events in SET1 and $P_{T}^{jet} > 40~ GeV$ and
7$\div$15 $\%$ of events in SET2 and $P_{T}^{jet} > 50~ GeV$ and .
The Z+jets background is less affected by $b$ jet cut,
60$\div$70 $\%$ of events  are retained. (see Fig.6)
\par
Since the Z+jets and ZW backgrounds do not contain top quark, applyment
of $t \rightarrow Zq$ mass cut is very useful for the rejection of Z+jets
and ZW backgrounds.
 In Fig. 5.b the distribution
of reconstructed invariant top mass $m_{llj}$ (in reconstruction $b$-jet
does not participate)
for all combinations of $llj$
for the signal is presented. The resolution $\sigma$ of $m_{llj}$  is $\sigma$=14.
The reconstruction of top mass have been required within three mass
windows: $m_{Zq}$ $\pm 8$ GeV (narrow cut), $m_{Zq}$ $\pm 12$ GeV ($\sim ~ \sigma$)
$m_{Zq}$ $\pm 24$ GeV ($\sim ~2\sigma$).
The top mass cut suppressed strongly Z+jets  and ZW backgrounds, in
$m_{Zq}$ $\pm 8$ GeV mass window Z+jets have been vanished.
\par
We suppose that new physics will be discovered if the signal
significance is $S/\sqrt{B}\geq 5$.
The $S/\sqrt{B}$ is computed as
 $S/\sqrt{B}=N_{S}/\sqrt{\Sigma N_{B}}$, where $\Sigma N_{B}$ is the sum over
the all observed backgrounds events, $N_{S}$ -- is the number of observed events
for the signal
\begin{center}
$N_{S}= N_{exp}^{S}  \cdot \epsilon_{l}  \cdot
\epsilon^{S}$
\end{center}
the lepton identification efficience is $\epsilon_{l}= 90\%$
and  $\epsilon^{S}$ is the signal efficiency.
\par
 The
Branching ratio for the t $\rightarrow$ Zq decay  is expressed by the following
formula:
\begin{eqnarray}
Br(t \rightarrow Zq)= K \cdot \sqrt{\Sigma N_{B}}/
 \epsilon^{S}  \label{eq4}
\end{eqnarray}
where K=2.52 $\cdot$ 10$^{-6}$.
The estimated  Branching ratios for the leptonic mode
for two  sets of kinematical cuts and
 $P_{T}^{jet} > 40~ GeV$ and $P_{T}^{jet} > 50~ GeV$
are presented  in Table 9. One can see from Table 9, that for
the estimation of Branching ratios more optimal is $P_{T}^{jet} > 50~ GeV$
cut. In each m$_{Zq}$ mass window the estimated Branching ratios for the both
SETs are less for $P_{T}^{jet} > 50~ GeV$ than for $P_{T}^{jet} > 40~ GeV$.
In this channel, a value of Br(t $\rightarrow$ Zq ) as low as 1.1x10$^{-4}$
could be discovered at the 5$\sigma$ level with an integrated luminosity
10$^{5}$ pb$^{-1}$.
\par
The Branching ratio sensitivity for $t \rightarrow Zq$ leptonic mode have been
estimated by Columbia University group [13] for a small statistic at
10$^{4}$ pb$^{-1}$ luminosity. Their results are in a good agreemnet with
our ones.
\   \par
\   \par
\   \par
\   \par
{\Large{\bf{2.The Hadronic Decay Mode}}}
\   \par
\   \par
\   \par
\par
The hadronic decay mode of the channel $t \rightarrow Zc$ had been studied
by a fast simulation package ATLFAST 2.14. The basic information of the detector
geometry is used by this package: the $\eta$ coverage for the calorimetry, the
size of the barrel/endcap transition region for the electromagnetic calorimeter
and the granularity of the calorimeters. No effects related for the detailed
shapes of particle showers in the calorimeters, for the charged track
multiplicity in jets, etc are taken into account. The main goal of the ATLFAST
package is to simulate and analyse fully generated events and select isolated leptons,
reconstructed jets, label $b$-jets, $c$-jets and estimate the missing transverse
energy. A more or less accurate parametrization of leptons momentum resolution
is included, as well as a parametrization of the hadronic calorimeter energy
resolution and the effect of the ATLAS magnetic field on jet reconstruction.
For the generation of a processes the ATLFAST uses the generator PYTHIA 5.7.
The routines of ATLFAST-B rundomly simulate $b,~c$ jets tagging and provide jet
energy recalibration. In our case the events have been simulated with ATLFAST
and analysed the results using the ATLFAST-B utilities whenever.
\par
For recalibration of light jets momenta we have considered two calibration
factors: 1) $K_{jet}=P_{T}^{q(q=c,u)}/P_{T}^{jet}$  2) $K_{jet}=P_{T}^{parton}/P_{T}^{jet}$
where $P_{T}^{parton}$ represents transverse momenta of light quarks originated
from $W \rightarrow qq$ decay. The first factor is used for top invariant mass
reconstruction $m_{llj}$ and the second one for $W$ invariant mass $m_{jj}$
calculation.
\par
 The $t\bar{t}$ events had been generated
at $\sqrt{s}=14$ TeV and $m_{top}=175$ GeV with proton structure
functions CTEQ2L. Initial and final state QED and QCD (ISR, FSR) radiation,
multiple interactions, fragmentations and decays of unstabled particles
were enabled as for leptonic mode.
\par
The signal
$t\bar{t} \rightarrow ZcWb \rightarrow l^{+} l^{-} jjjj_{b}$  signature
for hadronic mode is: two isolated charged leptons, which reconstruct
$Z$ and 4 energetic jets.
\par
This mode has the following backgrounds:\\
$\bullet$  $Z( \rightarrow ll) +jets$\\
$\bullet$ $pp \rightarrow W^{\pm}Z+X \rightarrow jj l^{+} l^{-}+X$\\
$\bullet$ $t\bar{t} \rightarrow W^{+}bW^{-} \tilde {b}$ final state topology:\\
a) $l^{+} \nu j_{b}l^{-} \tilde {\nu}j_{b}$ \\
b) $jjj_{b}jjj{b}$\\
c) $l^{\pm} \nu j_{b}jjj_{b}$\\
In case of a) decay additional two jets are QCD jets and in b) and c) items
the source of leptons is cascade decays. Therefore if we demand two isolated
energetic leptons and four jets among which only one $b$-jet is required,
$t\bar{t}$ background is significantly suppressed. A few survived events
vanish on the level of $m_{jj}$ and $m_{jjb}$ invariant masses reconstructions.
\par
To speed up the generation of Z+jets background the same thresholds as in
the case of the leptonic mode, has been imposed on invariant mass
$\hat{m}=\sqrt{\hat{s}}$
and transverse momentum $\hat{p_{\perp}}$.
\par
The following
preselection cuts were applied to signal and background before
writing the events on DST:\\
1. The event contains at least 2 leptons (electrons with $P_{T} > 5 $ GeV
within pseudorapidity $|\eta| < 2.5$ and muons with
$P_{T} > 6 $ GeV within pseudorapidity $|\eta| < 2.4$).\\
2. The events must contain a pair of opposite sign and same flavour
leptons compatible with them having come from Z bosons.\\
3. The number of jets in event $N_{jet} \geq 4$ with $P_{T}^{jet} > 15 $ GeV.
\par
 The expected number of events $N_{exp}$  for hadronic mode are:\\
$\bullet$ for $ZW$ background
\begin{center}
$N_{exp}^{ZW}= 121000$
\end{center}
\par
After preselection cuts the following kinematical cuts had been applied
in sequence for the signal and backgrounds:
\  \par
\  \par
\begin{tabular}{ll}
Lepton $P_{T}$ cut:& 2l with $P_{T}^{l} > 20~ GeV$ in $|\eta^{l}| < 2.5 $
for e$^{+}$,e$^{-}$ and\\
&in $|\eta^{l}| < 2.4$ for $\mu^{+}$, $\mu^{-}$\\
jet $P_{T}$ cut:& 4 jets with $P_{T}^{jet} > 50~ GeV$
in $|\eta^{l}| < 2.5 $\\
jets isolation cut:& ${\Delta}R_{jj} > 0.4 $\\
Lepton-jets isolation cut:& ${\Delta}R_{lj} > 0.4 $\\
Z mass cut:& $M_{Z} \pm 6~ GeV$\\
W mass cut:& $M_{W} \pm 16~ GeV$\\
$b$ jets cut:& only one tagged $b$ jet in the event\\
t$\rightarrow$W$^{+}$b mass cut:&$M_{Wb} \pm 8~ GeV$\\
t$\rightarrow$Zq mass cut:&$M_{Zq} \pm 8~ GeV$,
$M_{Zq} \pm 12~ GeV$,
$M_{Zq} \pm 24~ GeV$\\
\end{tabular}
\   \par
\   \par
The number of accepted events and efficiencies ($\%$) of each of these cuts
in sequence for signal and backgrounds are presented in Tables 10$\div$12.
After preselection cuts 46 $\%$ of events are accepted for signal, while
for Z+jets background -- 3.5 $\%$ and for ZW -- 4.1 $\%$. The requirement of two isolated leptons with
$P_{T}^{l} > 20~ GeV$ in $|\eta^{l}| < 2.5 $ does not affect the signal and
backgrounds. After this cut 80 -- 90 $\%$ of events are survived.
\par
The demand of four jets with with $P_{T}^{jet} > 50~ GeV$ in $|\eta^{l}| < 2.5 $
 strongly reduces Z+jets and ZW
backgrounds. In Z+jets  11 $\%$ of events accepted by the previous cut
are retained and in ZW -- 9 $\%$.  The jets isolation
and lepton-jets isolation cuts does not reduce the signal and backgrounds,
98 --99 $\%$ of events are retained.
\par
For the reconstruction of Z mass, have been required the same mass windows as
for the leptonic mode. Fig.7.a presents the distribution of reconstructed
invariant mass of $ll$ pairs $m_{ll}$ for the best combinations of $ll$ for
the signal. The resolution of the reconstructed Z mass is $\sigma$=2.91 GeV.
Thus mass window $\pm$6 GeV corresponds to $\sim ~2\sigma$. After applying Z mass
cut, in signal and Z+jets background 85 $\%$ of events accepted by the previous
cut are retained. In ZW background survived 74 $\%$ of events from those accepted
by the lepton-jets isolation cut.
\par
In hadronic mode the additional kinematical variable which can be used to seperate
signal from backgrounds is the reconstructed W mass.
For the
reconstruction of W mass, have been required a pair of jets within
$m_{jj} \pm 16~ GeV$ mass window. The jets that reconstruct the best
combinations of $W$ invariant mass, do not participate in further reconstructions.
Fig.8.a shows the distribution of reconstructed
invariant mass of jets pairs $m_{jj}$ for the best combinations of $jj$ for the
signal. The resolution $\sigma_{m_{jj}}$ = 8 GeV. Thus the mass window
$\pm$ 16 GeV corresponds to $2\sigma$ of the reconstructed W in the signal.
 The W mass cut strongly suppresses Z+jets background
(since there is no W in this background). Only 28 $\%$ of events survive after
W mass cut in this background.
\par
The requirement of only one tagged $b$-jet in the event does not affect the signal,
whereas suppresses the Z+jets (retained 6 $\%$ of events) and ZW
backgrounds (survived 8 $\%$ of events).
\par
The reconstructed top mass t$\rightarrow$Wb cut is used to reduce Z+jets and
ZW backgrounds. In Fig.8.b the distribution of the reconstructed invariant top
mass (t$\rightarrow$Wb) $m_{jjj_{b}}$ for the best combinations of $jjj_{b}$
for the
signal is presented. The resolution of $m_{jjj_{b}}$ is
$\sigma_{m_{jjj_{b}}}$ =18.5 GeV.
For the reconstruction of t$\rightarrow$Wb mass narrow cut window have been chosen
$m_{Wb} \pm 8~ GeV$. After applying the t$\rightarrow$Wb  mass cut in signal survived
25 $\%$ of events accepted by the $b$ - jet cut. This cut strongly
suppresses Z+jets
background, 6 $\%$ of events are retained after this cut and
 in ZW background accepts only one event.
\par
For the reconstruction of top mass
t$\rightarrow$Zq have been chosen the same mass windows as for the leptonic
mode:
 $m_{Zq}$ $\pm 8$ GeV (narrow cut), $m_{Zq}$ $\pm 12$ GeV ($\sim ~ \sigma$)
$m_{Zq}$ $\pm 24$ GeV ($\sim ~2\sigma$). After applying this cut to the signal,
in mass window $m_{Zq}$ $\pm 8$ GeV are retained 43 $\%$ of events accepted by
the t$\rightarrow$Wb cut, in $m_{Zq}$ $\pm 12$ GeV -- 55 $\%$ and in
$m_{Zq}$ $\pm 24$ GeV -- 70 $\%$. It is worth to mention, that in mass window
$m_{Zq}$ $\pm 8$ GeV 91 $\%$ of events are reconstructed by $c$ jets ,
t$\rightarrow llj_{c}$ and only 10 $\%$ of events are reconstructed by
light jets. By the widening the mass window
 the mixture of events
reconstructed by light jets increases from
10 $\%$ to 22 $\%$, but the number of events reconstructed by $c$-jets
increases too.
 The Z+jets background
in mass window $m_{Zq}$ $\pm 8$ GeV vanishes
 and one event is accepted in $m_{Zq}$ $\pm 12$ GeV  and two events in
$m_{Zq}$ $\pm 24$ GeV. This cut completely suppresses the ZW
background in all mass windows.
 In Fig.7.b the distribution of reconstructed  t$\rightarrow$Zq invariant
top mass $m_{llj}$ for the best combinations of $llj$
is presented for the signal. The resolution $\sigma$ of $m_{llj}$ distribution
is $\sigma_{m_{llj}}$ = 9.9 GeV.
\par
In Table 13 are presented  relative efficiences in
$\%$ of some important kinamatical cut for leptonic and hadronic modes of the
signal for the comparison. One can see, that the
relative efficiences of leptons, jets , Z mass , $b$ jet  and
top mass cuts for leptonic and hadronic modes coincide, as it was expected.
\par
The Branching ratios for the hadronic mode have been estimated by the formula
(4), where for hadronic mode K=6.87 $\cdot$ 10$^{-7}$. The results are presented
in Table 14. In hadronic mode,
a value of Br(t $\rightarrow$ Zq ) as low as 2.2 $\div$ 2.3x10$^{-4}$
could be discovered at the 5$\sigma$ level with an integrated luminosity
10$^{5}$ pb$^{-1}$.
\par
Combining the results from the leptonic and hadronic modes a Branching ratios
have been estimated for t$\rightarrow$Zq. For the estimation the following formula
was used:
\begin{eqnarray}
Br(t \rightarrow Zq)= K \cdot \sqrt{\Sigma N_{B}}/
(0.216 \cdot \epsilon^{L} + 0.676 \cdot \epsilon^{H})  \label{eq13}
\end{eqnarray}
where $\epsilon^{L}$ is the signal efficiency for the leptonic mode,
$\epsilon^{H}$ is the signal efficiency for the hadronic mode and
K=4.65 $\cdot$ 10$^{-6}$. The results for the combined leptonic and hadronic
modes are presented in Table 15. One can see from the Table, that a Branching
ratio for
t $\rightarrow$ Zq  as low as 0.9x10$^{-4}$
could be discovered at the 5$\sigma$ level with an integrated luminosity
10$^{5}$ pb$^{-1}$.
\newpage
{\Large{\bf{3. Conclusions}}}
\   \par
\   \par
\par
We have studied the sensitivity to the top quark rare decay
via flavour-changing neutral currents
  $t \rightarrow Zq$ ($q=u,c$) at LHC on the ATLAS experiment for the
luminosity $L= 10^{5}~pb^{-1}$.
The Branching ratios
of $t \rightarrow Zq$ in pure leptonic and hadronic decay modes had been estimated.
The results demonstrate, that in the leptonic mode a Branching ratio
 as low as 1.1x10$^{-4}$  and in the hadronic mode
as low as 2.2 $\div$ 2.3x10$^{-4}$
could be discovered at the 5$\sigma$ level with an integrated luminosity
10$^{5}$ pb$^{-1}$. Combining the results from the leptonic and
hadronic modes, a Branching ratio for $t \rightarrow Zq$ as low as
0.9x10$^{-4}$ could be discovered at the 5$\sigma$ level.
\   \par
\   \par
\   \par
ACKNOWLEDGEMENTS
\   \par
\   \par
\par
We are very indebted to D.Froudevaux, J.Parsons, M.Cobal,
E.Richter-Wass, S.Slabospitsky for very interesting and important discussions.
We are very grateful to R.Mehdiyev
for many valuable advices.
We would like to thank P.Jenni and T.Grigalashvili for their  continuous
support and encouragement during our work, J.Khubua for providing
the opportunity to proceed our work in future.
The authors wish to thank Z. Menteshashvili for helping during
the preparation of the article.
\newpage
\par

\newpage
\begin{center}
\bf{ FIGURE CAPTIONS  }
\end{center}
\  \par
\  \par
\  \par
\  \par
{\bf Fig.1.}
 The $P_{T}$ distributions of leptons that recostruct $Z$ mass ,for leptonic
mode signal and backgrounds, normalised to unity.
The arrow indicates the threshold value $P_{T}^{l}$ for kinematical cut.\\
\   \par
{\bf Fig.2.}
 The third lepton $P_{T}$ distributions for leptonic
mode signal and backgrounds, normalised to unity.
The arrow indicates the threshold value $P_{T}^{l}$ for kinematical cut.\\
\   \par
{\bf Fig.3.}
 The reconstructed $P_{T}^{miss}$ distributions
for leptonic
mode signal and backgrounds, normalised to unity.
The arrow indicates the threshold value $P_{T}^{miss}$ for kinematical cut.\\
\   \par
{\bf Fig.4.}
 Distribution of jet multiplicity (threshold at $P_{T}^{jet} > 50~ GeV$)
for signal and backgrounds, normalised to unity.
The arrow indicates the threshold value of the number of jets
 for kinematical cut.\\
\   \par
{\bf Fig.5.}
a) Distribution of reconstructed invariant mass of the lepton pairs,
$M_{ll}$ for the leptonic mode.
b) Distribution of reconstructed invariant mass of
$t \rightarrow llj$ for the leptonic mode.\\
\   \par
{\bf Fig.6.}
 Distribution of $b$-jet multiplicity (threshold at $P_{T}^{jet} > 50~ GeV$)
for signal and backgrounds, normalised to unity.\\
\   \par
{\bf Fig.7.}
a) Distribution of reconstructed invariant mass of the lepton pairs,
$M_{ll}$ for the best combination (hadronic mode).
b) Distribution of reconstructed invariant mass of
$t \rightarrow llj$ for the best combination of $llj$ (hadronic mode).\\
\   \par
{\bf Fig.8.}
a) Distribution of reconstructed invariant mass of the jet pairs,
$M_{jj}$ for the best combination (hadronic mode).
b) Distribution of reconstructed invariant mass of
$t \rightarrow jjj_b$ for the best combination of $jjj_b$ (hadronic mode).\\
\newpage
Table 1. The numbers of events and efficiencies ($\%$) of kinematical cuts
applied in
\par
sequence for the signal (leptonic mode) for $P_T^{jet} > 40 $ GeV
\   \par
\   \par
\begin {tabular}{|l|c|c|c|c|}
\hline
\hline
\multicolumn{1}{|c|}{ C U T S}&\multicolumn{2}{|c|}{S E T 1 }&
\multicolumn{2}{|c|}{S E T 2 }\\
\cline{2-5}
\multicolumn{1}{|c|}{}&\multicolumn{1}{|c|}{Events}&
\multicolumn{1}{|c|}{Effic. ($\%$)}&
\multicolumn{1}{|c|}{Events}&\multicolumn{1}{|c|}{Effic. ($\%$)}\\
\hline
\hline
Number of generated events&20565&&20565&\\
\hline
Preselection cuts&16497&80.2&16497&80.2\\
\hline
Lepton isolation cut:&     &&&\\
No track with $P_{T} > 2~ GeV$ in cone
&14071&68.4&12672&61.6\\
 ${\Delta}R=0.2$ (SET1); ${\Delta}R=0.3$ (SET2) &&&&\\
\hline
3l with $P_{T}^{l} > 20~ GeV$ in $|\eta^{l}| < 2.5 $
&9844&47.9&8885&43.3\\
\hline
$P_{T}^{miss} > 30 ~GeV$
&7462&36.3& 6730&32.7\\
\hline
2 jets with $|\eta^{jet}| < 2.5 $
&7021&34.1&6307&30.7\\
\hline
2 jets with $P_{T}^{jet} > 40~ GeV$
&6213&30.2&5552&27.0\\
\hline
jets isolation: ${\Delta}R_{jj} > 0.4 $
&6179&30.1&5522&26.8\\
\hline
Lepton-jets isolation: ${\Delta}R_{lj} > 0.4 $
&5573&27.1&5138&24.9\\
\hline
\hline
Z mass $M_{Z} \pm 6~ GeV$
& 4717&22.9&4347&22.1\\
\hline
one tagged $b$ jet in the event
& 2367&11.5&2193&10.7\\
\hline
t$\rightarrow$Zq mass : $M_{Zq} \pm 8~ GeV$&1038&5.1&970&4.7\\
\hline
t$\rightarrow$Zq mass : $M_{Zq} \pm 12~ GeV$&1389&6.8&1294&6.3\\
\hline
t$\rightarrow$Zq mass : $M_{Zq} \pm 24~ GeV$&1879&9.1&1742&8.5\\
\hline
\end{tabular}
\newpage
Table 2. The numbers of events and efficiencies ($\%$) of kinematical cuts
applied in
\par
sequence for the signal (leptonic mode) for $P_T^{jet} > 50 $ GeV
\   \par
\   \par
\begin {tabular}{|l|c|c|c|c|}
\hline
\hline
\multicolumn{1}{|c|}{ C U T S}&\multicolumn{2}{|c|}{S E T 1 }&
\multicolumn{2}{|c|}{S E T 2 }\\
\cline{2-5}
\multicolumn{1}{|c|}{}&\multicolumn{1}{|c|}{Events}&
\multicolumn{1}{|c|}{Effic. ($\%$)}&
\multicolumn{1}{|c|}{Events}&\multicolumn{1}{|c|}{Effic. ($\%$)}\\
\hline
\hline
Number of generated events&20565&&20565&\\
\hline
Preselection cuts&16497&80.2&16497&80.2\\
\hline
Lepton isolation cut:&     &&&\\
No track with $P_{T} > 2~ GeV$ in cone
&14071&68.4&12672&61.6\\
 ${\Delta}R=0.2$ (SET1); ${\Delta}R=0.3$ (SET2) &&&&\\
\hline
3l with $P_{T}^{l} > 20~ GeV$ in $|\eta^{l}| < 2.5 $
&9844&47.9&8885&43.3\\
\hline
$P_{T}^{miss} > 30 ~GeV$
&7462&36.3& 6730&32.7\\
\hline
2 jets with $|\eta^{jet}| < 2.5 $
&7021&34.1&6307&30.7\\
\hline
2 jets with $P_{T}^{jet} > 50~ GeV$
&4904&23.8&4341&21.1\\
\hline
jets isolation: ${\Delta}R_{jj} > 0.4 $
&4881&23.7&4322&21.0\\
\hline
Lepton-jets isolation: ${\Delta}R_{lj} > 0.4 $
&4453&21.6&4063&19.8\\
\hline
\hline
Z mass $M_{Z} \pm 6~ GeV$
& 3781&18.4&3450&16.8\\
\hline
one tagged $b$ jet in the event
& 1835&8.9&1678&8.2\\
\hline
t$\rightarrow$Zq mass : $M_{Zq} \pm 8~ GeV$&787&3.8&728&3.5\\
\hline
t$\rightarrow$Zq mass : $M_{Zq} \pm 12~ GeV$&1053&5.1&973&4.7\\
\hline
t$\rightarrow$Zq mass : $M_{Zq} \pm 24~ GeV$&1383&6.7&1264&6.1\\
\hline
\end{tabular}
\newpage
Table 3. The numbers of events and efficiencies ($\%$) of kinematical cuts
applied in
\par
sequence for the Z+jets background (leptonic mode) for $P_T^{jet} > 40 $ GeV
\   \par
\   \par
\begin {tabular}{|l|c|c|c|c|}
\hline
\hline
\multicolumn{1}{|c|}{ C U T S}&\multicolumn{2}{|c|}{S E T 1 }&
\multicolumn{2}{|c|}{S E T 2 }\\
\cline{2-5}
\multicolumn{1}{|c|}{}&\multicolumn{1}{|c|}{Events}&
\multicolumn{1}{|c|}{Effic. ($\%$)}&
\multicolumn{1}{|c|}{Events}&\multicolumn{1}{|c|}{Effic. ($\%$)}\\
\hline
\hline
Number of generated events&21.4$\cdot 10^{6}$&&21.4$\cdot 10^{6}$&\\
\hline
Preselection cuts&366843&1.7&366843&1.7\\
\hline
Lepton isolation cut:&     &&&\\
No track with $P_{T} > 2~ GeV$ in cone
&40472&0.19&21539&0.10\\
 ${\Delta}R=0.2$ (SET1); ${\Delta}R=0.3$ (SET2) &&&&\\
\hline
3l with $P_{T}^{l} > 20~ GeV$ in $|\eta^{l}| < 2.5 $
&1644&7.7$\cdot 10^{-3}$&945&4.4$\cdot 10^{-3}$\\
\hline
$P_{T}^{miss} > 30 ~GeV$
&129&6.0$\cdot 10^{-4}$& 80&3.7$\cdot 10^{-4}$\\
\hline
2 jets with $|\eta^{jet}| < 2.5 $
&122&5.7$\cdot 10^{-4}$&80&3.7$\cdot 10^{-4}$\\
\hline
2 jets with $P_{T}^{jet} > 40~ GeV$
&97&4.5$\cdot 10^{-4}$&56&2.6$\cdot 10^{-4}$\\
\hline
jets isolation: ${\Delta}R_{jj} > 0.4 $
&94&4.4$\cdot 10^{-4}$&56&2.6$\cdot 10^{-4}$\\
\hline
Lepton-jets isolation: ${\Delta}R_{lj} > 0.4 $
&59&2.8$\cdot 10^{-4}$&38&1.8$\cdot 10^{-4}$\\
\hline
\hline
Z mass $M_{Z} \pm 6~ GeV$
& 45&2.1$\cdot 10^{-4}$&28&1.3$\cdot 10^{-4}$\\
\hline
one tagged $b$ jet in the event
& 28&1.3$\cdot 10^{-4}$&21&9.8$\cdot 10^{-5}$\\
\hline
t$\rightarrow$Zq mass : $M_{Zq} \pm 8~ GeV$&0&0&0&0\\
\hline
t$\rightarrow$Zq mass : $M_{Zq} \pm 12~ GeV$&1&4.8$\cdot 10^{-6}$&1&4.8$\cdot 10^{-6}$\\
\hline
t$\rightarrow$Zq mass : $M_{Zq} \pm 24~ GeV$&2&9.3$\cdot 10^{-6}$&1&4.8$\cdot 10^{-6}$\\
\hline
\end{tabular}

\newpage
Table 4. The numbers of events and efficiencies ($\%$) of kinematical cuts
applied in
\par
sequence for the Z+jets background (leptonic mode) for $P_T^{jet} > 50 $ GeV
\   \par
\   \par
\begin {tabular}{|l|c|c|c|c|}
\hline
\hline
\multicolumn{1}{|c|}{ C U T S}&\multicolumn{2}{|c|}{S E T 1 }&
\multicolumn{2}{|c|}{S E T 2 }\\
\cline{2-5}
\multicolumn{1}{|c|}{}&\multicolumn{1}{|c|}{Events}&
\multicolumn{1}{|c|}{Effic. ($\%$)}&
\multicolumn{1}{|c|}{Events}&\multicolumn{1}{|c|}{Effic. ($\%$)}\\
\hline
\hline
Number of generated events&21.4$\cdot 10^{6}$&&21.4$\cdot 10^{6}$&\\
\hline
Preselection cuts&366843&1.7&366843&1.7\\
\hline
Lepton isolation cut:&     &&&\\
No track with $P_{T} > 2~ GeV$ in cone &40472&0.19&21539&0.10\\
 ${\Delta}R=0.2$ (SET1); ${\Delta}R=0.3$ (SET2) &&&&\\
\hline
3l with $P_{T}^{l} > 20~ GeV$ in $|\eta^{l}| < 2.5 $
&1644&7.7$\cdot 10^{-3}$&945&4.4$\cdot 10^{-3}$\\
\hline
$P_{T}^{miss} > 30 ~GeV$
&129&6.0$\cdot 10^{-4}$& 80&3.7$\cdot 10^{-4}$\\
\hline
2 jets with $|\eta^{jet}| < 2.5 $
&122&5.7$\cdot 10^{-4}$&80&3.7$\cdot 10^{-4}$\\
\hline
2 jets with $P_{T}^{jet} > 50~ GeV$
&63&2.9$\cdot 10^{-4}$&45&2.1$\cdot 10^{-4}$\\
\hline
jets isolation: ${\Delta}R_{jj} > 0.4 $&63&2.9$\cdot 10^{-4}$&45&2.1$\cdot 10^{-4}$\\
\hline
Lepton-jets isolation: ${\Delta}R_{lj} > 0.4 $&42&1.9$\cdot 10^{-4}$&31&1.5$\cdot 10^{-4}$\\
\hline
\hline
Z mass $M_{Z} \pm 6~ GeV$& 35&1.6$\cdot 10^{-4}$&24&1.1$\cdot 10^{-4}$\\
\hline
one tagged $b$ jet in the event& 24&1.1$\cdot 10^{-4}$&10&4.7$\cdot 10^{-5}$\\
\hline
t$\rightarrow$Zq mass : $M_{Zq} \pm 8~ GeV$&0&0&0&0\\
\hline
t$\rightarrow$Zq mass : $M_{Zq} \pm 12~ GeV$&0&0&0&0\\
\hline
t$\rightarrow$Zq mass : $M_{Zq} \pm 24~ GeV$&1&4.8$\cdot 10^{-6}$&0&0 \\
\hline
\end{tabular}
\newpage
Table 5. The numbers of events and efficiencies ($\%$) of kinematical cuts
applied in
\par
sequence for the Z+W background (leptonic mode) for $P_T^{jet} > 40 $ GeV
\   \par
\   \par
\begin {tabular}{|l|c|c|c|c|}
\hline
\hline
\multicolumn{1}{|c|}{ C U T S }&\multicolumn{2}{|c|}{S E T 1 }&
\multicolumn{2}{|c|}{S E T 2 }\\
\cline{2-5}
\multicolumn{1}{|c|}{}&\multicolumn{1}{|c|}{Events}&
\multicolumn{1}{|c|}{Effic. ($\%$)}&
\multicolumn{1}{|c|}{Events}&\multicolumn{1}{|c|}{Effic. ($\%$)}\\
\hline
\hline
Number of generated events&35700&&35700&\\
\hline
Preselection cuts&2941&8.2&2941&8.2\\
\hline
Lepton isolation cut:&     &&&\\
No track with $P_{T} > 2~ GeV$ in cone
& 2842&7.9&2630&7.4\\
 ${\Delta}R=0.2$ (SET1); ${\Delta}R=0.3$ (SET2) &&&&\\
\hline
3l with $P_{T}^{l} > 20~ GeV$ in $|\eta^{l}| < 2.5 $
& 1920&5.4&1778&5.0\\
\hline
$P_{T}^{miss} > 30 ~GeV$
&1326&3.7& 1252&3.5\\
\hline
2 jets with $|\eta^{jet}| < 2.5 $
& 997&2.8&938&2.6\\
\hline
2 jets with $P_{T}^{jet} > 40~ GeV$
&539&1.5&509&1.4\\
\hline
jets isolation: ${\Delta}R_{jj} > 0.4 $
&532&1.5&502&1.4\\
\hline
Lepton-jets isolation: ${\Delta}R_{lj} > 0.4 $
&532&1.5&502&1.4\\
\hline
\hline
Z mass $M_{Z} \pm 6~ GeV$
& 442&1.2&421&1.2\\
\hline
one tagged $b$ jet in the event
& 32&0.09&28&0.08\\
\hline
t$\rightarrow$Zq mass : $M_{Zq} \pm 8~ GeV$&5&1.4$\cdot 10^{-2}$&5&1.4$\cdot 10^{-2}$\\
\hline
t$\rightarrow$Zq mass : $M_{Zq} \pm 12~ GeV$&7&2.0$\cdot 10^{-2}$&6&1.7$\cdot 10^{-2}$\\
\hline
t$\rightarrow$Zq mass : $M_{Zq} \pm 24~ GeV$&8&2.2$\cdot 10^{-2}$&7&2.0$\cdot 10^{-2}$\\
\hline
\end{tabular}
\newpage
Table 6. The numbers of events and efficiencies ($\%$) of kinematical cuts
applied in
\par
sequence for the Z+W background (leptonic mode) for $P_T^{jet} > 50 $ GeV
\   \par
\   \par
\begin {tabular}{|l|c|c|c|c|}
\hline
\hline
\multicolumn{1}{|c|}{ C U T S}&\multicolumn{2}{|c|}{S E T 1 }&
\multicolumn{2}{|c|}{S E T 2 }\\
\cline{2-5}
\multicolumn{1}{|c|}{}&\multicolumn{1}{|c|}{Events}&
\multicolumn{1}{|c|}{Effic. ($\%$)}&
\multicolumn{1}{|c|}{Events}&\multicolumn{1}{|c|}{Effic. ($\%$)}\\
\hline
\hline
Number of generated events&35700&&35700&\\
\hline
Preselection cuts&2941&8.2&2941&8.2\\
\hline
Lepton isolation cut:&     &&&\\
No track with $P_{T} > 2~ GeV$ in cone
& 2842&7.9&2630&7.4\\
 ${\Delta}R=0.2$ (SET1); ${\Delta}R=0.3$ (SET2) &&&&\\
\hline
3l with $P_{T}^{l} > 20~ GeV$ in $|\eta^{l}| < 2.5 $
& 1920&5.4&1778&5.0\\
\hline
$P_{T}^{miss} > 30 ~GeV$
&1326&3.7& 1252&3.5\\
\hline
2 jets with $|\eta^{jet}| < 2.5 $
& 997&2.8&938&2.6\\
\hline
2 jets with $P_{T}^{jet} > 50~ GeV$
&278&0.8&227&0.6\\
\hline
jets isolation: ${\Delta}R_{jj} > 0.4 $
&276&0.8&225&0.6\\
\hline
Lepton-jets isolation: ${\Delta}R_{lj} > 0.4 $
&276&0.8&225&0.6\\
\hline
\hline
Z mass $M_{Z} \pm 6~ GeV$
& 230&0.6&180&0.5\\
\hline
one tagged $b$ jet in the event
& 19&0.05&28&0.04\\
\hline
t$\rightarrow$Zq mass : $M_{Zq} \pm 8~ GeV$&1&2.8$\cdot 10^{-3}$&1&2.8$\cdot 10^{-3}$\\
\hline
t$\rightarrow$Zq mass : $M_{Zq} \pm 12~ GeV$&1&2.8$\cdot 10^{-3}$&1&2.8$\cdot 10^{-3}$\\
\hline
t$\rightarrow$Zq mass : $M_{Zq} \pm 24~ GeV$&2&5.6$\cdot 10^{-3}$&2&5.6$\cdot 10^{-3}$\\
\hline
\end{tabular}
\newpage
Table 7. The numbers of events and efficiencies ($\%$) of kinematical cuts
applied in
\par
sequence for the $t\bar{t}$ background (leptonic mode) for $P_T^{jet} > 40 $ GeV
\   \par
\   \par
\begin {tabular}{|l|c|c|c|c|}
\hline
\hline
\multicolumn{1}{|c|}{ C U T S  }&\multicolumn{2}{|c|}{S E T 1 }&
\multicolumn{2}{|c|}{S E T 2 }\\
\cline{2-5}
\multicolumn{1}{|c|}{}&\multicolumn{1}{|c|}{Events}&
\multicolumn{1}{|c|}{Effic. ($\%$)}&
\multicolumn{1}{|c|}{Events}&\multicolumn{1}{|c|}{Effic. ($\%$)}\\
\hline
\hline
Number of generated events&3.72$\cdot 10^{6}$&&3.72$\cdot 10^{6}$&\\
\hline
Preselection cuts&848160&29.4&848160&29.4\\
\hline
Lepton isolation cut:&     &&&\\
No track with $P_{T} > 2~ GeV$ in cone
&93000&2.5&39432&1.1\\
 ${\Delta}R=0.2$ (SET1); ${\Delta}R=0.3$ (SET2) &&&&\\
\hline
3l with $P_{T}^{l} > 20~ GeV$ in $|\eta^{l}| < 2.5 $
&3378&9.1$\cdot 10^{-2}$&1858&5.0$\cdot 10^{-2}$\\
\hline
$P_{T}^{miss} > 30 ~GeV$
&3036&8.2$\cdot 10^{-2}$&1600&4.3$\cdot 10^{-2}$\\
\hline
2 jets with $|\eta^{jet}| < 2.5 $
&2902&7.8$\cdot 10^{-2}$&1339&3.6$\cdot 10^{-2}$\\
\hline
2 jets with $P_{T}^{jet} > 40~ GeV$
&2604&3.6$\cdot 10^{-2}$&1116&2.0$\cdot 10^{-2}$\\
\hline
jets isolation: ${\Delta}R_{jj} > 0.4 $
&2604&3.6$\cdot 10^{-2}$&1116&2.0$\cdot 10^{-2}$\\
\hline
Lepton-jets isolation: ${\Delta}R_{lj} > 0.4 $
&1488&4.0$\cdot 10^{-2}$&744&1.4$\cdot 10^{-2}$\\
\hline
\hline
Z mass $M_{Z} \pm 6~ GeV$
&186&5.0$\cdot 10^{-3}$&50&1.3$\cdot 10^{-3}$\\
\hline
one tagged $b$ jet in the event
& 74&2.0$\cdot 10^{-3}$&20&5.4$\cdot 10^{-4}$\\
\hline
t$\rightarrow$Zq mass : $M_{Zq} \pm 8~ GeV$&10&2.7$\cdot 10^{-4}$&10&2.7$\cdot 10^{-4}$\\
\hline
t$\rightarrow$Zq mass : $M_{Zq} \pm 12~ GeV$&12&3.2$\cdot 10^{-4}$&11&3.0$\cdot 10^{-4}$\\
\hline
t$\rightarrow$Zq mass : $M_{Zq} \pm 24~ GeV$&14&3.8$\cdot 10^{-4}$&13&3.5$\cdot 10^{-4}$\\
\hline
\end{tabular}
\newpage
Table 8. The numbers of events and efficiencies ($\%$) of kinematical cuts
applied in
\par
sequence for the $t\bar{t}$ background (leptonic mode) for $P_T^{jet} > 50 $ GeV
\   \par
\   \par
\begin {tabular}{|l|c|c|c|c|}
\hline
\hline
\multicolumn{1}{|c|}{ C U T S}&\multicolumn{2}{|c|}{S E T 1 }&
\multicolumn{2}{|c|}{S E T 2 }\\
\cline{2-5}
\multicolumn{1}{|c|}{}&\multicolumn{1}{|c|}{Events}&
\multicolumn{1}{|c|}{Effic. ($\%$)}&
\multicolumn{1}{|c|}{Events}&\multicolumn{1}{|c|}{Effic. ($\%$)}\\
\hline
\hline
Number of generated events&3.72$\cdot 10^{6}$&&3.72$\cdot 10^{6}$&\\
\hline
Preselection cuts&848160&29.4&848160&29.4\\
\hline
Lepton isolation cut:&     &&&\\
No track with $P_{T} > 2~ GeV$ in cone
&93000&2.5&39432&1.1\\
 ${\Delta}R=0.2$ (SET1); ${\Delta}R=0.3$ (SET2) &&&&\\
\hline
3l with $P_{T}^{l} > 20~ GeV$ in $|\eta^{l}| < 2.5 $
&3378&9.1$\cdot 10^{-2}$&1858&5.0$\cdot 10^{-2}$\\
\hline
$P_{T}^{miss} > 30 ~GeV$
&3036&8.2$\cdot 10^{-2}$&1600&4.3$\cdot 10^{-2}$\\
\hline
2 jets with $|\eta^{jet}| < 2.5 $
&2902&7.8$\cdot 10^{-2}$&1339&3.6$\cdot 10^{-2}$\\
\hline
2 jets with $P_{T}^{jet} > 50~ GeV$
&1339&3.6$\cdot 10^{-2}$&740&2.0$\cdot 10^{-2}$\\
\hline
jets isolation: ${\Delta}R_{jj} > 0.4 $
&1332&3.6$\cdot 10^{-2}$& 736&2.0$\cdot 10^{-2}$\\
\hline
Lepton-jets isolation: ${\Delta}R_{lj} > 0.4 $
&1012&2.7$\cdot 10^{-24}$&596&1.6$\cdot 10^{-2}$\\
\hline
\hline
Z mass $M_{Z} \pm 6~ GeV$
&104&2.8$\cdot 10^{-3}$&29&7.8$\cdot 10^{-4}$\\
\hline
one tagged $b$ jet in the event
& 45&1.2$\cdot 10^{-3}$&10&2.7$\cdot 10^{-4}$\\
\hline
t$\rightarrow$Zq mass : $M_{Zq} \pm 8~ GeV$& 3&8.1$\cdot 10^{-5}$& 3&8.1$\cdot 10^{-5}$\\
\hline
t$\rightarrow$Zq mass : $M_{Zq} \pm 12~ GeV$&5&1.3$\cdot 10^{-4}$& 4&1.1$\cdot 10^{-4}$\\
\hline
t$\rightarrow$Zq mass : $M_{Zq} \pm 24~ GeV$&6&1.6$\cdot 10^{-4}$& 5&1.3$\cdot 10^{-4}$\\
\hline
\end{tabular}
\newpage
Table 9. The estimated Branching ratios
 $t \rightarrow Zq$ ($q=c,u$) for leptonic mode
\   \par
\   \par
\   \par
\   \par
\   \par
\begin {tabular}{|l|l|l|c|c|}
\hline
\hline
\multicolumn{3}{|l|}{ }&\multicolumn{1}{|c|}{S E T 1 }&
\multicolumn{1}{|c|}{S E T 2 }\\
\hline
\hline
&  & && \\
& & $m_{Zq}\pm 8$ GeV&1.91$\cdot 10^{-4}$ &2.07$\cdot 10^{-4}$\\
&  & && \\
\cline{3-5}
&  & && \\
&$P_T^{jet} > 40 $ GeV & $m_{Zq}\pm 12$  GeV&1.65$\cdot 10^{-4}$ &1.69$\cdot 10^{-4}$\\
&  & && \\
\cline{3-5}
                    && && \\
 & & $m_{Zq}\pm 24$ GeV&1.35$\cdot 10^{-4}$
&1.36$\cdot 10^{-4}$\\
Br(t$\rightarrow$Zq)&  & && \\
\cline{2-5}
&  & && \\
&& $m_{Zq}\pm 8$ GeV&1.32$\cdot 10^{-4}$ &
1.44$\cdot 10^{-4}$\\
&  & && \\
\cline{3-5}
&  & && \\
&$P_T^{jet} > 50 $ GeV & $m_{Zq}\pm 12$  GeV&1.21$\cdot 10^{-4}$ &1.20$\cdot 10^{-4}$\\
&  & && \\
\cline{3-5}
&  & && \\
 & & $m_{Zq}\pm 24$ GeV&1.12$\cdot 10^{-4}$
&1.11$\cdot 10^{-4}$\\
&  & && \\
\hline
\end{tabular}
\newpage
Table 10. The numbers of events and efficiencies ($\%$) of kinematical cuts
applied in
\par
sequence for the signal (hadronic mode).
\   \par
\   \par
\begin {tabular}{|l|c|c|}
\hline
\hline
\multicolumn{1}{|c|}{ C U T S}&\multicolumn{1}{|c|}{Events}&
\multicolumn{1}{|c|}{Effic. ($\%$)}\\
\hline
\hline
Number of generated events&\multicolumn{1}{|c|}{19002}&
\multicolumn{1}{|c|}{}\\
\hline
Preselection cuts&\multicolumn{1}{|c|}{8742}&\multicolumn{1}{|c|}{46.0}\\
\hline
2l with $P_{T}^{l} > 20~ GeV$ in $|\eta^{l}| < 2.5 $
&\multicolumn{1}{|c|}{7174}&\multicolumn{1}{|c|}{37.7}\\
\hline
\hline
4 jets with $P_T > 50 ~GeV$ in $|\eta^{jet}| < 2.5$&2896 &15.2\\
\hline
jets isolation: ${\Delta}R_{jj} > 0.4 $
&2828   &14.9\\
\hline
Lepton-jets isolation: ${\Delta}R_{lj} > 0.4 $ &2826  &14.9\\
\hline
Z mass $M_{Z} \pm 6~ GeV$
&2426&12.8 \\
\hline
W mass $M_{W} \pm 16~ GeV$
&1006 & 5.3\\
\hline
one tagged $b$ jet in the event
&432&2.2\\
\hline
t$\rightarrow$Wb mass: $M_{Wb} \pm 8~ GeV$
&106&0.6\\
\hline
t$\rightarrow$Zq mass : $M_{Zq} \pm 8~ GeV$&46
&0.2\\
\hline
t$\rightarrow$Zq mass : $M_{Zq} \pm 12~ GeV$&58
&0.3\\
\hline
t$\rightarrow$Zq mass : $M_{Zq} \pm 24~ GeV$
 &74&0.4\\
\hline
\end{tabular}
\newpage
Table 11. The numbers of events and efficiencies ($\%$) of kinematical cuts
applied in
\par
sequence for the Z+jets background (hadronic mode).
\   \par
\   \par
\begin {tabular}{|l|c|c|}
\hline
\hline
\multicolumn{1}{|c|}{ C U T S}&\multicolumn{1}{|c|}{Events}&
\multicolumn{1}{|c|}{Effic. ($\%$)}\\
\hline
\hline
Number of generated events&\multicolumn{1}{|c|}{21.4$\cdot 10^{6}$}&
\multicolumn{1}{|c|}{}\\
\hline
Preselection cuts&\multicolumn{1}{|c|}{746259}&\multicolumn{1}{|c|}{3.5}\\
\hline
2l with $P_{T}^{l} > 20~ GeV$ in $|\eta^{l}| < 2.5 $
&\multicolumn{1}{|c|}{592627}&\multicolumn{1}{|c|}{2.8}\\
\hline
\hline
4 jets with $P_T > 50~ GeV$ in $|\eta^{jet}| < 2.5$&63478 &0.29\\
\hline
jets isolation: ${\Delta}R_{jj} > 0.4 $
&60421   &0.28\\
\hline
Lepton-jets isolation: ${\Delta}R_{lj} > 0.4 $
&60394   &0.28\\
\hline
Z mass $M_{Z} \pm 6~ GeV$
&50973&0.24\\
\hline
W mass $M_{W} \pm 16~ GeV$
&14170&6.6$\cdot 10^{-2}$\\
\hline
one tagged $b$ jet in the event
&1379&6.4$\cdot 10^{-3}$\\
\hline
t$\rightarrow$Wb mass: $M_{Wb} \pm 8~ GeV$
&90&4.2$\cdot 10^{-4}$\\
\hline
t$\rightarrow$Zq mass : $M_{Zq} \pm 8~ GeV$&0
&0\\
\hline
t$\rightarrow$Zq mass : $M_{Zq} \pm 12~ GeV$&1
&4.8$\cdot 10^{-6}$\\
\hline
t$\rightarrow$Zq mass : $M_{Zq} \pm 24~ GeV$
 &2&9.3$\cdot 10^{-6}$\\
\hline
\end{tabular}
\newpage
Table 12. The numbers of events and efficiencies ($\%$) of kinematical cuts
applied in
\par
sequence for the  Z+W background (hadronic mode).
\   \par
\   \par
\begin {tabular}{|l|c|c|}
\hline
\hline
\multicolumn{1}{|c|}{ C U T S}&\multicolumn{1}{|c|}{Events}&
\multicolumn{1}{|c|}{Effic. ($\%$)}\\
\hline
\hline
Number of generated events&\multicolumn{1}{|c|}{121000}&
\multicolumn{1}{|c|}{}\\
\hline
Preselection cuts&\multicolumn{1}{|c|}{4970}&\multicolumn{1}{|c|}{4.1}\\
\hline
2l with $P_{T}^{l} > 20~ GeV$ in $|\eta^{l}| < 2.5 $
&\multicolumn{1}{|c|}{4456}&\multicolumn{1}{|c|}{3.7}\\
\hline
\hline
4 jets with $P_T > 50~ GeV$ in $|\eta^{jet}| < 2.5$&400 &0.3\\
\hline
jets isolation: ${\Delta}R_{jj} > 0.4 $
       &390   & 0.3\\
\hline
Lepton-jets isolation: ${\Delta}R_{lj} > 0.4 $
       &361  & 0.3\\
\hline
Z mass $M_{Z} \pm 6~ GeV$
    &268& 0.2 \\
\hline
W mass $M_{W} \pm 16~ GeV$
    &139 & 0.1\\
\hline
one tagged $b$ jet in the event
   &11&9.1$\cdot 10^{-3}$\\
\hline
t$\rightarrow$Wb mass: $M_{Wb} \pm 8~ GeV$
&1&8.3$\cdot 10^{-4}$\\
\hline
t$\rightarrow$Zq mass : $M_{Zq} \pm 8~ GeV$ &0
&0\\
\hline
t$\rightarrow$Zq mass : $M_{Zq} \pm 12~ GeV$ &0
&0\\
\hline
t$\rightarrow$Zq mass : $M_{Zq} \pm 24~ GeV$
 &0 &0\\
\hline
\end{tabular}
\newpage
Table 13. The  relative efficiencies ($\%$)
of some kinematical cuts
\par
for leptonic and hadronic modes of the signal.
\   \par
\   \par
\begin {tabular}{|l|c|c|}
\hline
\hline
\multicolumn{1}{|c|}{ C U T S}&\multicolumn{1}{|c|}{LEPTONIC MODE}&
\multicolumn{1}{|c|}{HADRONIC MODE}\\
\cline{2-3}
\multicolumn{1}{|c|}{}&
\multicolumn{1}{|c|}{Rel.Effic. ($\%$)}&
\multicolumn{1}{|c|}{Rel.Effic. ($\%$)}\\
\hline
\hline
3l with $P_{T}^{l} > 20~ GeV$ in $|\eta^{l}| < 2.5 $
&70.1&  \\
\hline
2l with $P_{T}^{l} > 20~ GeV$ in $|\eta^{l}| < 2.5 $
 &&82.0 \\
\hline
2 isolated jets with $P_{T}^{jet} > 50~ GeV$
&48.6&   \\
\hline
4 isolated jets with $P_{T}^{jet} > 50~ GeV$
& &39.4 \\
\hline
Lepton-jets isolation: ${\Delta}R_{lj} > 0.4 $
&94.0&99.0\\
\hline
\hline
Z mass $M_{Z} \pm 6~ GeV$
&84.9&85.8\\
\hline
one tagged $b$ jet in the event
&48.6&42.9\\
\hline
t$\rightarrow$Zq mass : $M_{Zq} \pm 8~ GeV$&43.4&43.4\\
\hline
t$\rightarrow$Zq mass : $M_{Zq} \pm 12~ GeV$&49.2&54.7\\
\hline
t$\rightarrow$Zq mass : $M_{Zq} \pm 24~ GeV$&75.3&69.8\\
\hline
\end{tabular}
\newpage
Table 14. The estimated Branching ratios
 $t \rightarrow Zq$ ($q=c,u$) for
 for hadronic mode.
\   \par
\   \par
\   \par
\   \par
\   \par
\begin {tabular}{|l|l|l|c|}
\hline
\hline
\multicolumn{3}{|l}{ }&\multicolumn{1}{c|}{ }\\
\hline
\hline
                    &  & & \\
 & & $m_{Zq}\pm 8$ GeV&2.86$\cdot 10^{-4}$ \\
&  & & \\
\cline{3-4}
&  & & \\
Br(t$\rightarrow$Zq)&$P_T^{jet} > 50 $ GeV  & $m_{Zq}\pm 12$  GeV&2.20$\cdot 10^{-4}$ \\
& && \\
\cline{3-4}
& && \\
& & $m_{Zq}\pm 24$ GeV&2.30$\cdot 10^{-4}$\\
& && \\
\hline
\end{tabular}
\newpage
Table 15. The estimated Branching ratios
 $t \rightarrow Zq$ ($q=c,u$)
 for combined leptonic and hadronic modes.
\   \par
\   \par
\   \par
\   \par
\   \par
\begin {tabular}{|l|l|l|c|c|}
\hline
\hline
\multicolumn{3}{|l|}{ }&\multicolumn{1}{|c|}{S E T 1 }&
\multicolumn{1}{|c|}{S E T 2 }\\
\hline
\hline
&  & && \\
& & $m_{Zq}\pm 8$  GeV&0.95$\cdot 10^{-4}$ &1.01$\cdot 10^{-4}$\\
&  & && \\
\cline{3-5}
&  & && \\
Br(t$\rightarrow$Zq)&$P_T^{jet} > 50 $ GeV & $m_{Zq}\pm 12$  GeV&0.94$\cdot 10^{-4}$ &0.93$\cdot 10^{-4}$\\
&  & && \\
\cline{3-5}
&  & && \\
 & & $m_{Zq}\pm 24$ GeV&0.90$\cdot 10^{-4}$
&0.88$\cdot 10^{-4}$\\
&  & && \\
\hline
\end{tabular}
\newpage
\begin{figure}
\begin{center}
\epsfig{file=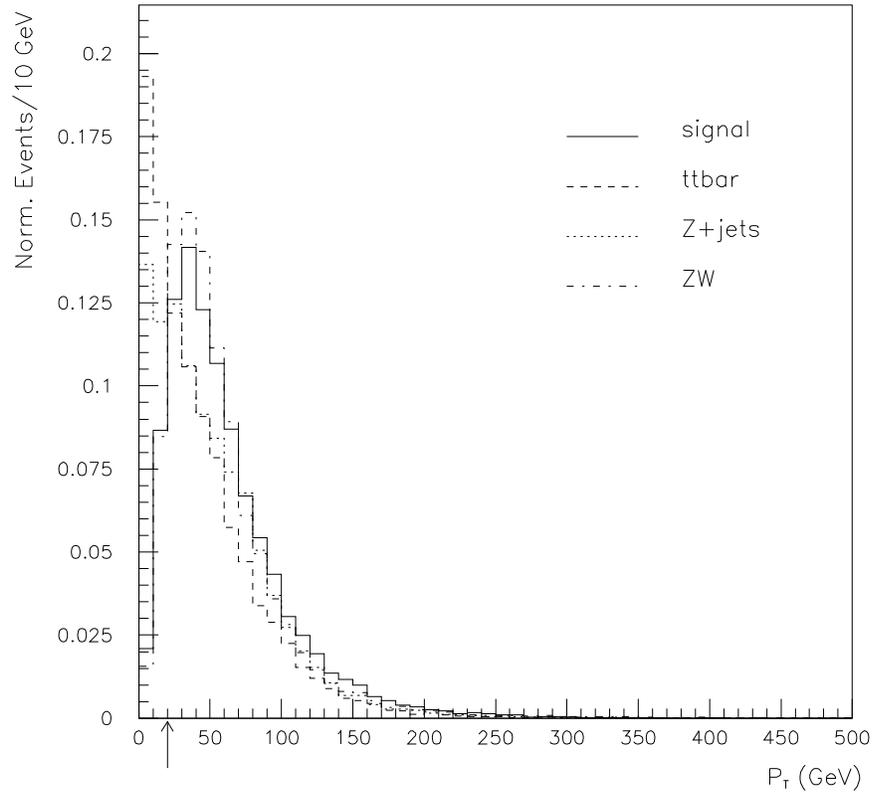,bbllx=0pt,bblly=0pt,bburx=594pt,bbury=842pt,
width=18cm,angle=0}
\end{center}
\vspace{-6.4cm}
\hspace{0.cm}
\begin{minipage}{16.0cm}
\caption
{ The $P_{T}$ distributions of leptons that recostruct $Z$ mass ,for leptonic
mode signal and backgrounds, normalised to unity.
The arrow indicates the threshold value $P_{T}^{l}$ for kinematical cut.}
\end{minipage}
\end{figure}
\newpage
\begin{figure}
\begin{center}
\epsfig{file=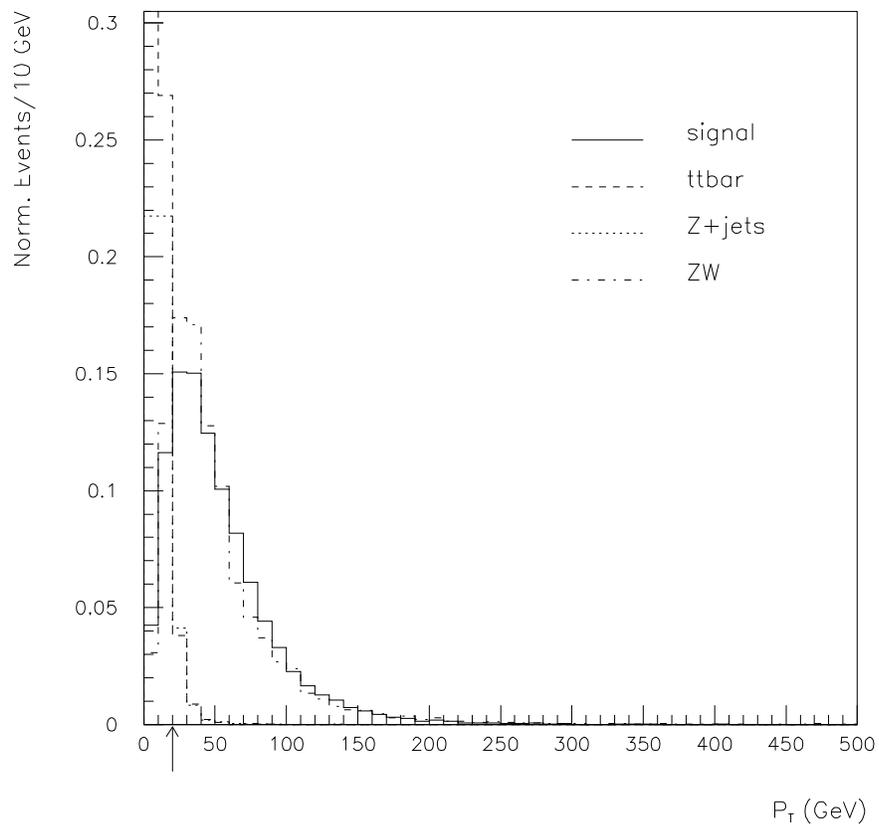,bbllx=0pt,bblly=0pt,bburx=594pt,bbury=842pt,
width=18.0cm,angle=0}
\end{center}
\vspace{-7.cm}
\hspace{0.cm}
\begin{minipage}{16.0cm}
\caption
{ The third lepton $P_{T}$ distributions for leptonic
mode signal and backgrounds, normalised to unity.
The arrow indicates the threshold value $P_{T}^{l}$ for kinematical cut.}
\end{minipage}
\end{figure}
\newpage
\begin{figure}
\begin{center}
\epsfig{file=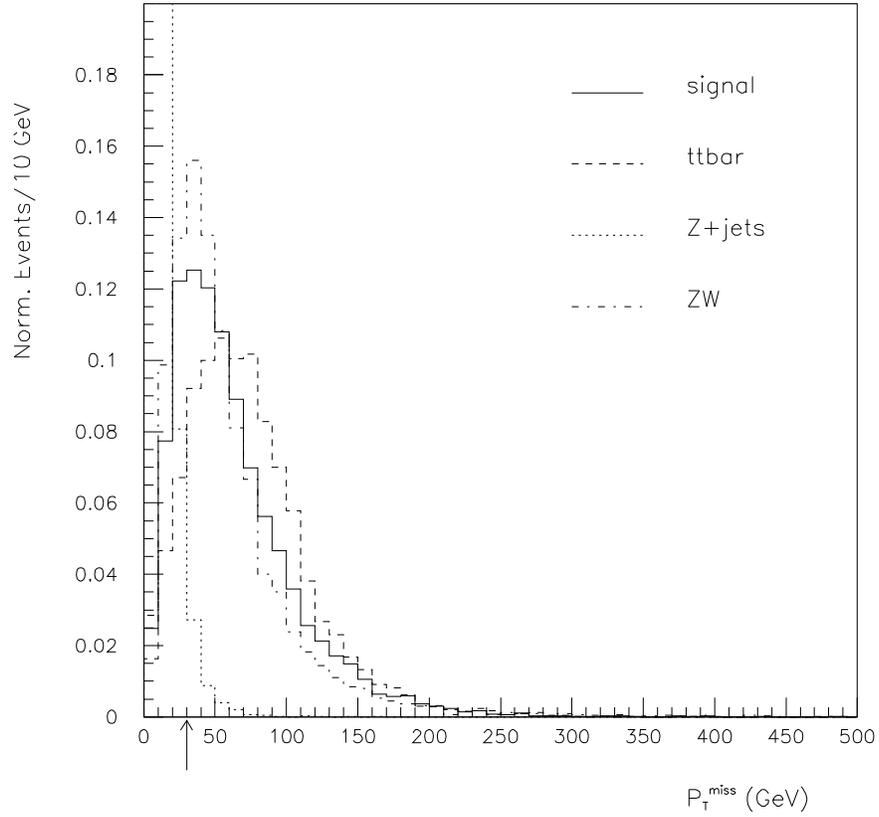,bbllx=0pt,bblly=0pt,bburx=594pt,bbury=842pt,
width=18cm,angle=0}
\end{center}
\vspace{-6.4cm}
\hspace{0.cm}
\begin{minipage}{16.0cm}
\caption
{ The reconstructed $P_{T}^{miss}$ distributions
for leptonic
mode signal and backgrounds, normalised to unity.
The arrow indicates the threshold value $P_{T}^{miss}$ for kinematical cut.}
\end{minipage}
\end{figure}
\newpage
\begin{figure}
\begin{center}
\epsfig{file=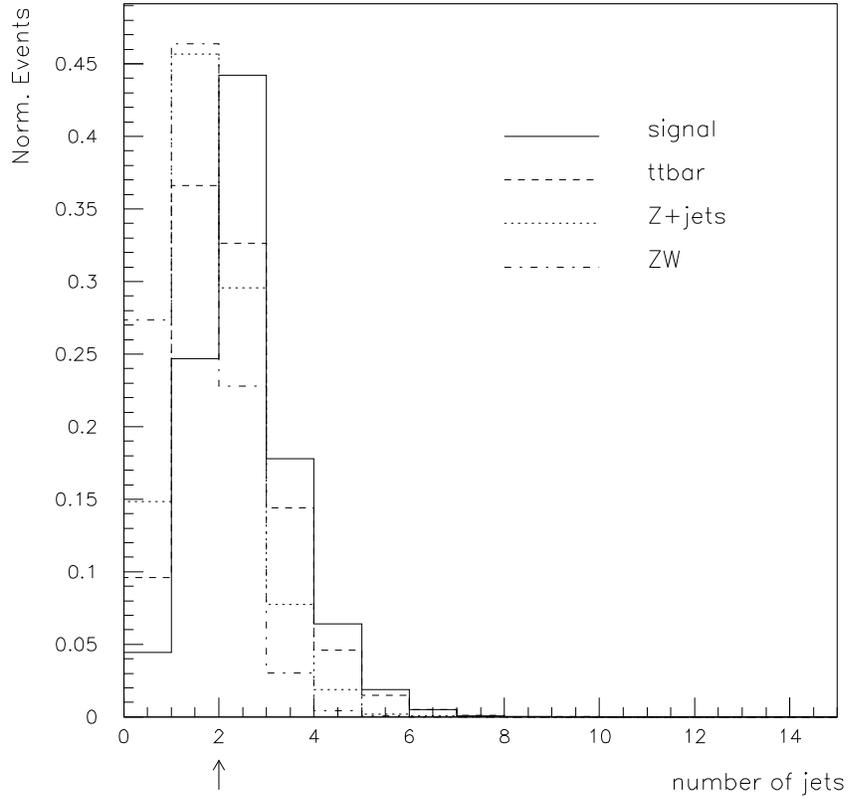,bbllx=0pt,bblly=0pt,bburx=594pt,bbury=842pt,
width=18cm,angle=0}
\end{center}
\vspace{-6.4cm}
\hspace{0.cm}
\begin{minipage}{16.0cm}
\caption
{ Distribution of jet multiplicity (threshold at $P_{T}^{jet} > 50~ GeV$)
for signal and backgrounds, normalised to unity.
The arrow indicates the threshold value of the number of jets
 for kinematical cut.}
\end{minipage}
\end{figure}
\newpage
\begin{figure}
\begin{center}
\epsfig{file=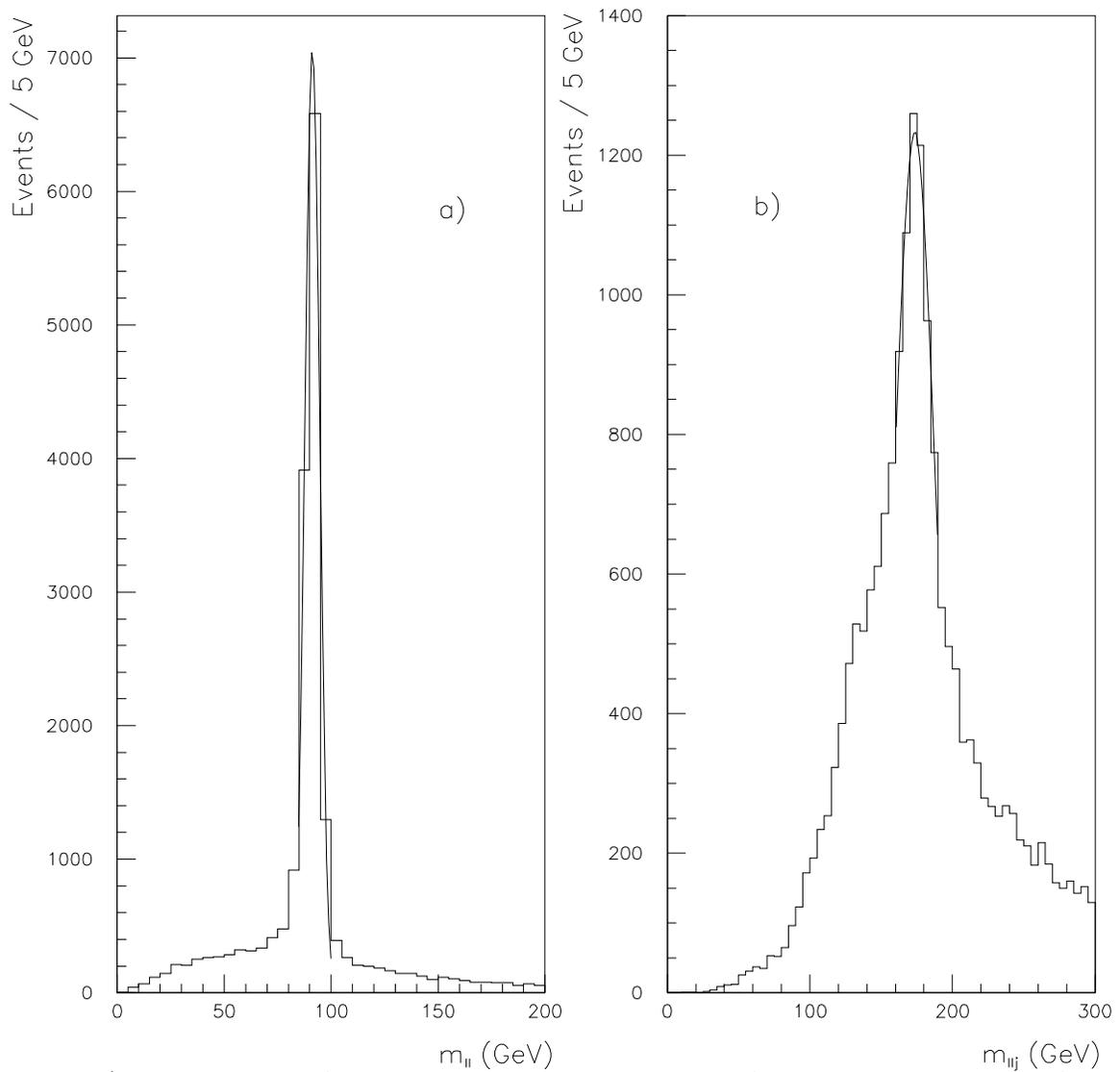,bbllx=0pt,bblly=0pt,bburx=594pt,bbury=842pt,
width=18cm,angle=0}
\end{center}
\vspace{-5.8cm}
\hspace{0.cm}
\begin{minipage}{18.0cm}
\caption
{a) Distribution of reconstructed invariant mass of the lepton pairs,
$M_{ll}$ for the leptonic mode.
b) Distribution of reconstructed invariant mass of
$t \rightarrow llj$ for the leptonic mode.}
\end{minipage}
\end{figure}
\newpage
\begin{figure}
\begin{center}
\epsfig{file=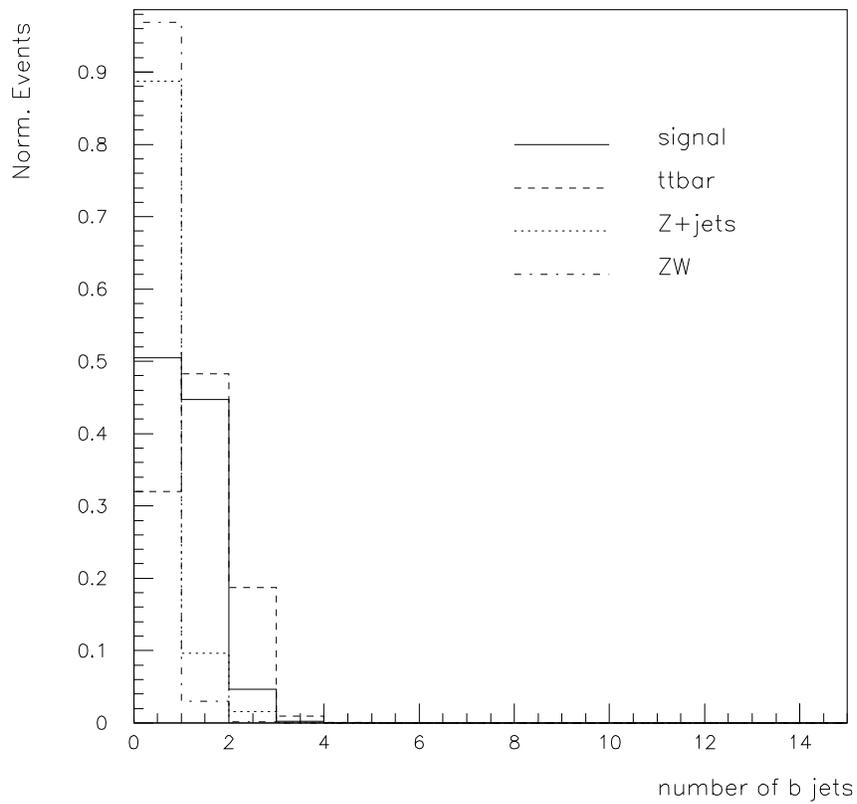,bbllx=0pt,bblly=0pt,bburx=594pt,bbury=842pt,
width=18cm,angle=0}
\end{center}
\vspace{-6.7cm}
\hspace{0.cm}
\begin{minipage}{16.0cm}
\caption
{ Distribution of $b$-jet multiplicity (threshold at $P_{T}^{jet} > 50~ GeV$)
for signal and backgrounds, normalised to unity.}
\end{minipage}
\end{figure}
\newpage
\begin{figure}
\begin{center}
\epsfig{file=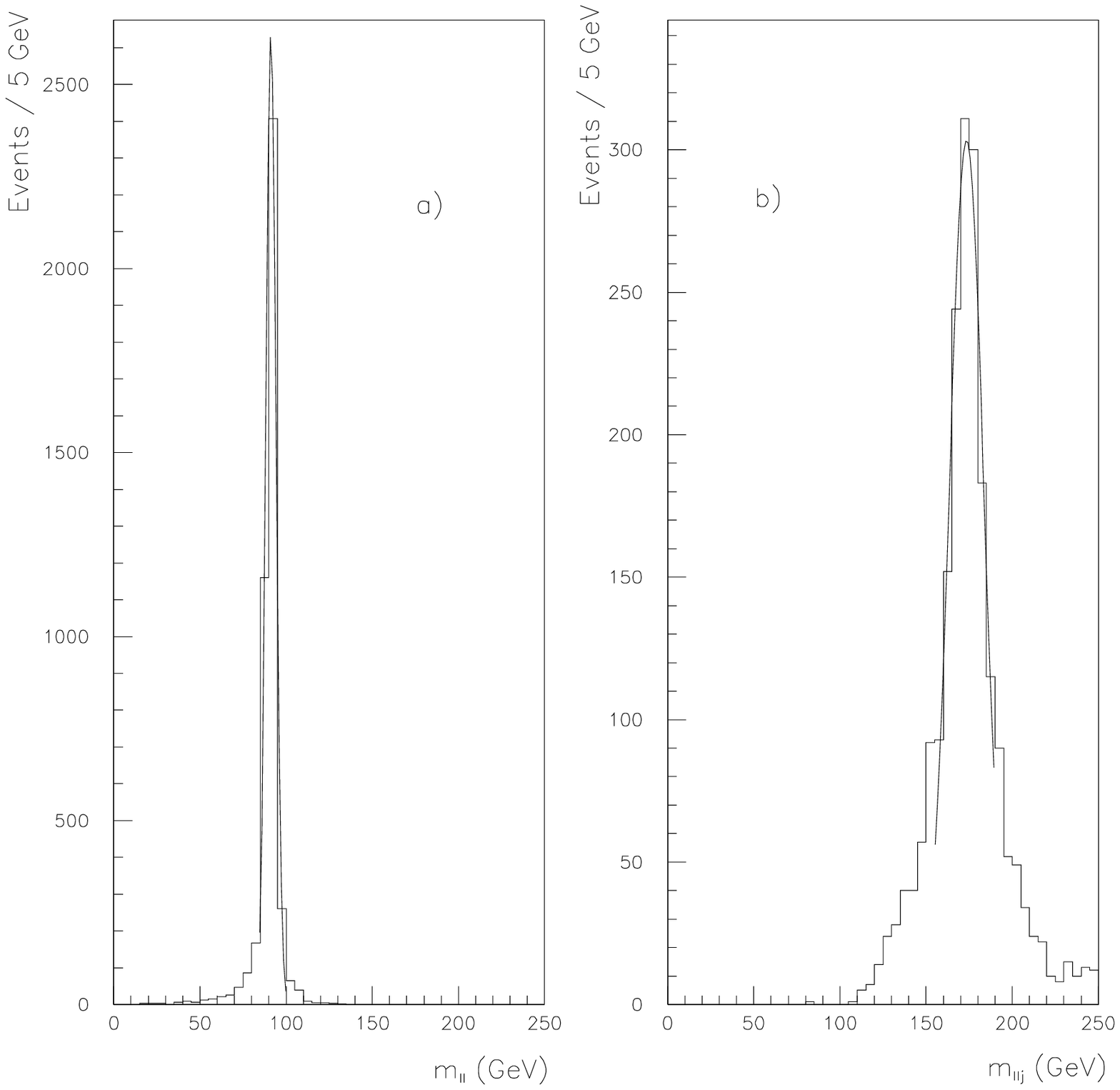,bbllx=0pt,bblly=0pt,bburx=594pt,bbury=842pt,
width=18cm,angle=0}
\end{center}
\vspace{-5.8cm}
\hspace{0.cm}
\begin{minipage}{18.0cm}
\caption
{a) Distribution of reconstructed invariant mass of the lepton pairs,
$M_{ll}$ for the best combination (hadronic mode).
b) Distribution of reconstructed invariant mass of
$t \rightarrow llj$ for the best combination of $llj$ (hadronic mode).}
\end{minipage}
\end{figure}
\newpage
\begin{figure}
\begin{center}
\epsfig{file=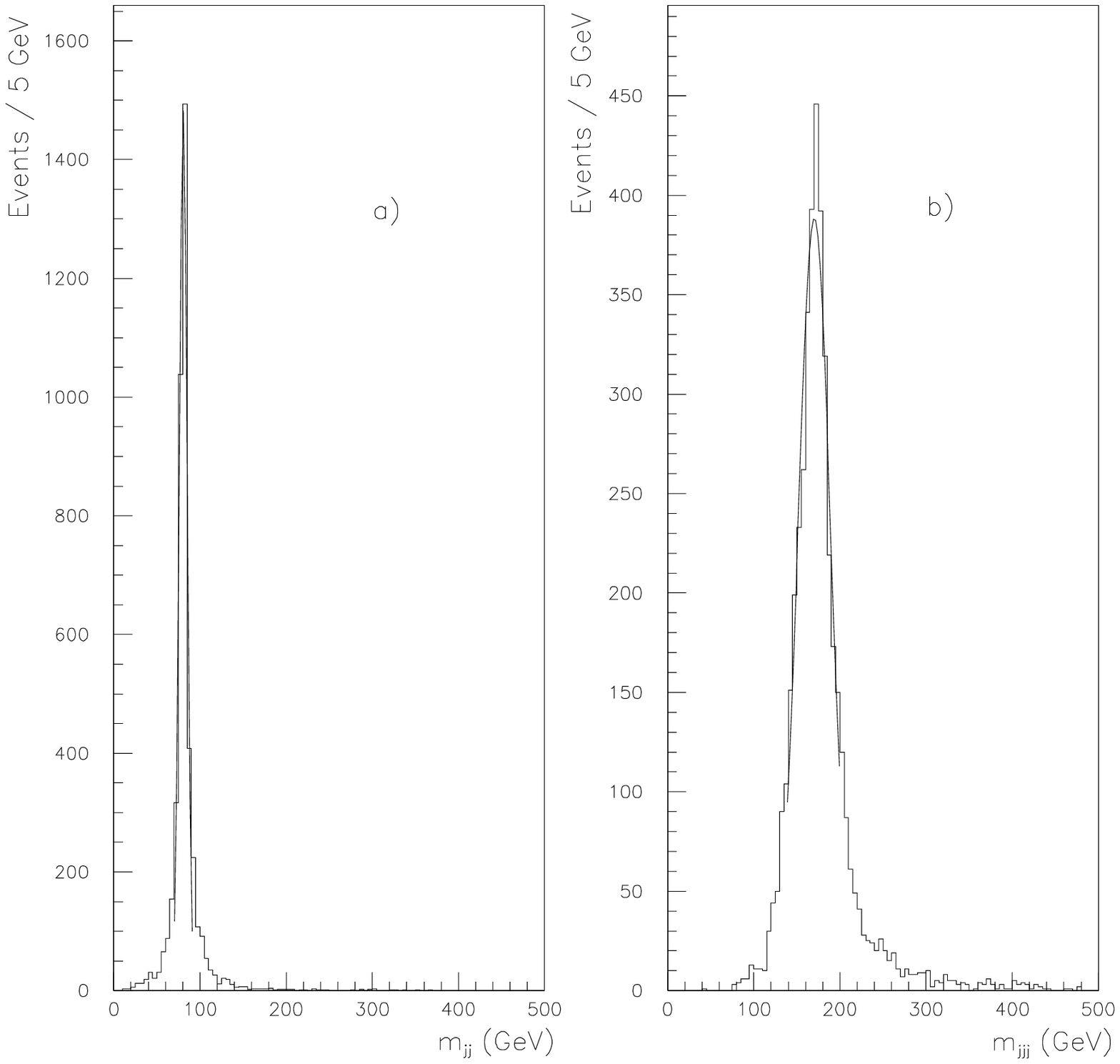,bbllx=0pt,bblly=0pt,bburx=594pt,bbury=842pt,
width=18cm,angle=0}
\end{center}
\vspace{-5.8cm}
\hspace{0.cm}
\begin{minipage}{18.0cm}
\caption
{a) Distribution of reconstructed invariant mass of the jet pairs,
$M_{jj}$ for the best combination (hadronic mode).
b) Distribution of reconstructed invariant mass of
$t \rightarrow jjj_b$ for the best combination of $jjj_b$ (hadronic mode).}
\end{minipage}
\end{figure}
\end{document}